\documentclass[letter,twocolumn,showpacs,preprintnumbers,amsmath,amssymb,nofootinbib]{revtex4}
\usepackage[english]{babel}
\usepackage{graphicx}
\usepackage{dcolumn}
\usepackage{bm}
\usepackage{verbatim}
\newcommand{\ket}[1]{\left| {#1} \right\rangle}

\newcommand{\braket}[2]{\left\langle {#1}\left|{#2}\right.\right\rangle}

\def\slashchar#1{\setbox0=\hbox{$#1$} 
\dimen0=\wd0 
\setbox1=\hbox{/} \dimen1=\wd1 
\ifdim\dimen0>\dimen1 
\rlap{\hbox to \dimen0{\hfil/\hfil}} 
#1 
\else 
\rlap{\hbox to \dimen1{\hfil$#1$\hfil}} 
/ 
\fi}

\begin{document}

\title{Physical qubits from charged particles: Infrared divergences in quantum information}

\author{Juan Le\'on}
\email{leon@imaff.cfmac.csic.es}
\homepage{http://www.imaff.csic.es/pcc/QUINFOG/}

\author{Eduardo Mart\'{i}n-Mart\'{i}nez}%
 \email{martin@imaff.cfmac.csic.es}
\homepage{http://www.imaff.csic.es/pcc/QUINFOG/}
\affiliation{Instituto de F\'{i}sica Fundamental, CSIC\\
Serrano 113-B, 28006 Madrid, Spain.\\
}%

\date{20 Jan 2009}

\begin{abstract}
We consider soft photons effects (IR structure of QED) on the construction of physical qubits. Soft-photons appear when we build charged qubits from the asymptotic states of QED. This construction is necessary in order to include the effect of soft photons on entanglement measures. The nonexistence of free charged particles (due to the long range of QED interactions) lead us to question the sense of the very concept of free charged qubit. In this letter, using the ``dressing'' formalism, we build physical charged qubits from dressed fields which have the correct asymptotic behavior, are gauge invariant, their propagators have a particle pole structure and are free from infrared divergences. Finally, we discuss the impact of the soft corrections on the entanglement measures.
\end{abstract}

\maketitle

Relativistic quantum information is quite a novel area of research whose scope covers from the effect of Lorentz transformations on entanglement measures to the information content of black holes \cite{Beckman,peresterno2,RevPeresTerno,reznik,fuentesschuller,Verch,AlsingSchul,LamataPRL,Caban,reznik2,Chinos,JleonSabin2a,JleonSabin2b}. Included among its main tasks should be the suitable treatment of the divergencies that plague quantum field theories. In this work we address the infrared problem in the construction of charged qubits.

The infrared divergences of QED are intimately connected to the fact that the very concept of free charged particle is alien to the theory \cite{KulishFaddeev,JR,BaganC}. Since the asymptotic limit of the QED Hamiltonian is not the free one, residual Coulomb-like interactions remain for $t \rightarrow \infty$. So, the EM interactions of charged particles never switch off and, as a consequence, they are always surrounded by a soft photon cloud whose inescapable presence has been argued in quantum information against the conception of free charged qubits. It has been used to say that the physical realization of a single charged qubit is itself an idealization \cite{RevPeresTerno} and that it should be upgraded with all the multi-soft-photon components.

We will show below that if we try to build the charged qubits taking into account the asymptotic interaction, another problem will come up: the states that evolve with the asymptotic Hamiltonian are not gauge invariant and, hence, they cannot be conceived as a physical entity. In addition, for these states we no longer obtain a pole-like contribution in the energy spectrum of the charged particles associated with their masses \cite{Infraparticles}.

Seemingly, all these arguments indicate that we cannot speak about physical charged qubits in QED. However, there is a method of recovering physical states from asymptotic QED, it is the dressing formalism \cite{BaganC} which we will use to build physical charged qubits overcoming all the above problems.

The dressing formalism restores the gauge invariance of the asymptotic states preserving their dynamics. The dressed fields, that are surrounded by soft photon clouds, turn to be asymptotically well-behaved, their propagators having a proper pole structure and --crucial for our purposes-- the S-matrix elements constructed in terms of dressed fields are IR-finite \cite{BaganPR}. In this letter we build dressed two qubits states that are physical and have all the desired properties we mentioned.

Due to the effect of soft photon clouds that inescapably surround the charged particles even at asymptotic times, the standard interaction picture, splitting the Hamiltonian into free and interaction Hamiltonians, works poorly. In fact, a finite interaction relic ${H}_{\text{int}}^{\text{as}}(t)=-e \int d^3 \bm x J_\mu^{\text{as}}(t,\bm x)\,A^\mu(t,\bm x)$ still remains for $t\rightarrow\infty$. The asymptotic current turns out to be
\begin{equation}\label{corr}J^{\text{as}}_\mu(t,x)=\int d^3p\frac{p_\mu}{p_0}\rho(\bm p)\delta^3\left(\bm x-\frac{\bm p}{p_0}t\right),\end{equation}
where $e \rho(\bm p\,)= e \sum_n \left[b_n^\dagger (\bm p) b_n(\bm p)-d_n^\dagger(\bm p)d_n(\bm p)\right]$. It describes the charge operator moving along the classical particle trajectory.

Strictly speaking, only zero momentum photons can be emitted or absorbed by ${H}_{\text{int}}^{\text{as}}$, but neglecting them in the asymptotic evolution, as done in standard perturbation theory, is the origin of the spurious infrared
divergencies that plague QED \cite{Chung,KulishFaddeev}. These can be cured by building a new picture in which we consider the complete asymptotic Hamiltonian ${H}^{\text{as}}(t) = H_0 + {H}_{\text{int}}^{\text{as}}(t)$, instead of only $H_0$, as the unperturbed Hamiltonian. In this new picture, the vector potential $A_\mu^{\text{as}}$ is simply the sum of the free field and the Coulombic field generated by the asymptotic current.

The  charged fields in this asymptotic picture are related to the free fields by the transformation \begin{equation}\label{UT}U(t)=\exp\left[R(t)\right]\exp\left[i\Phi(t)\right]\end{equation}
where the distortion operator $\exp\left[R(t)\right]$ is given by \cite{KulishFaddeev}
\begin{equation}\label{3}R(t)=e\int \frac{d^3 p\,d^3 k}{\sqrt{2k_0(2\pi)^3}}\frac{p^\mu}{pk}\rho(\bm p)\left[a_\mu^\dagger(\bm k)e^{i\frac{k\cdot p}{p_0}t}-\text{h.c.}\right]\end{equation}
with $a_\mu^\dagger$ as the photon creator operator, and the so-called phase operator by
\begin{equation}\label{4}\Phi(t)=\frac{e^2}{8\pi}\int\! :\rho(\bm p)\rho(\bm q):\frac{p\!\cdot\!q\,\operatorname{sign}(t)}{\sqrt{(pq)^2-m^4}}\ln\frac{|t|}{{t'}} d^3 p d^3 q\end{equation}

The perturbation theory built using this new picture does not present infrared divergences \cite{KulishFaddeev,Chung} but at the price of introducing new interactions --those that subtract the zero momentum photons--, and of replacing the free states with  asymptotic ones. We treat this point  considering a bipartite free state of two would-be qubits of the form
\[\ket\Psi=b_{\sigma_2}^\dagger(p_2)b_{\sigma_1}^\dagger(p_1)\ket0\]
where the $b_{\sigma}^\dagger(p)$ are fermion creator operators. The asymptotic picture version of this state can be obtained by applying to these operators the $U(t)$ transformation defined in \eqref{UT}, which gives
\begin{equation}\label{estadoas}
\ket\psi_{\text{as}}=e^{-ie^2\phi(u_r,t)}e^{W(p_1,p_2,t)}b^\dagger_{\sigma_{_1}}(p_1)b^\dagger_{\sigma_{_2}}(p_2)\ket0
\end{equation}
where the terms
\begin{equation}\label{vdoble}
W(p_1,p_2,t)=\frac{e}{(2\pi)^{\frac32}}\int\frac{d^3k}{\sqrt{2k_0}}\sum_{j=1}^2\frac{p_j^\mu}{p_j k}e^{i\frac{k p_j}{{p_j}_{_0}}t}a_\mu^\dagger(k)-\text{h.c.}\end{equation}
\begin{equation}\label{divfase}\phi=\frac{1}{4\pi}u^{-1}_r(p_1,p_2)\log\frac{|t|}{{t'}}\end{equation}
are the net effect of applying $R(t)$ and $\Phi(t)$ operators on the considered state, and where  $u_r(p_1,p_2)$ is the absolute value of the relative velocity between the two particles, $u(p,q)=\sqrt{1-m^4/(pq)^2}$.

We now run into the serious obstructions to this construction. Due to the presence of photon operators in $W(p_1,p_2,t)$, the asymptotic states are no longer gauge invariant and they do not present the pole-like structure associated with massive particles. Therefore, they are not good starting points to build physical qubits. This can be overcome by the introduction of a dressing operator $h(x)$ to restore the gauge invariance \cite{BaganC}.  The ``dressed field", defined as the product of the charged field $\psi$ and the dressing,
\begin{equation}\label{6b}\psi^\text{as}_{\text{d}}(x)=h^{-1}(x)\psi^{\text{as}}(x)\end{equation}
should satisfy two conditions; i) Gauge invariance and, ii) conservation of asymptotic dynamics. Given that for asymptotic times the dynamics is governed by soft photons --for which the charged particles appear as (infinitely) heavy-- the dressing has to preserve the heavy particle dynamics. For infinite mass fields, the charge's 4-velocity
$u^\mu$ is superselected \cite{Georgi} and the equation of motion for the field is
\begin{equation}\label{geor}u\cdot D\psi^{\text{as}}(x)=0\end{equation}
when the  matter field is minimally coupled. For the gauge invariant dressed field this translates into
\begin{equation}\label{8}u\cdot \partial\psi^{\text{as}}_{\text{d}}(x)=0,\;\,\text{i.e. }\,
u\cdot\partial h^{-1}(x)=-ie\, h^{-1}(x)\, u\cdot A^{\text{as}}(x).\end{equation}

As shown in \cite{BaganC}, the dressing operator $h_i^{-1}$ for the i-th particle consists of a distortion term and a phase term $h_i^{-1}(x)=e^{\chi_i(x)}e^{-iK_i(x)}$ --like $U(t)$ given by Eq. \eqref{UT}.-- In  the large $t$ limit of interest for asymptotic states,
\begin{equation}\label{dressingxi}\chi_i(x)= e \int\frac{d^3k}{(2\pi)^3}\frac{1}{\sqrt{2k_0}}\left(\frac{V_i^\mu
a_\mu(k)}{V_i\!\cdot\! k}e^{-ik\cdot x}-\text{H.c.}\right)\end{equation}
where $V_i^\mu=(\eta+v_i)^\mu(\eta-v_i)\cdot k-k^\mu$.
 $\eta$ is an unitary temporal vector $\eta=(1,\bm
0)$, and $v_i=(0,\bm v_i)$ where $\bm v_i$ is the 3-velocity of i-th the particle.
The phase operator can be written as
\begin{equation}\label{dressingphase}K_i(x)=e\int_{{t'}}^{t} (\eta+ v_i)^\mu\frac{\partial^\nu
F_{\nu\mu}}{\mathcal{G}\cdot\partial}\left[x(s)\right]ds\end{equation}
with the integral taken along the world line of the  massive particle parameterized as $x^\mu(s)=x^\mu+(s-x^0)(\eta+v)^\mu$. The operator ${1}/({\mathcal{G}\cdot\partial})$ is defined by its action \cite{BaganC} as
${1}/({\mathcal{G}\cdot\partial})f(\bm x)\equiv \int d^3 z G(\bm x-\bm z)f(\bm z)$
where
\begin{equation}\label{Ge}G(\bm x)=-\frac{1}{4\pi}\frac{\gamma}{\sqrt{\bm x\,^2+\gamma^2 (\bm v \cdot \bm x)^2}}.\end{equation}

The dressing procedure introduce new vertices whose Feynman rules obtained from $e^{\chi_i(x)} \psi(x)_{as}$ \cite{BaganPR} are given in Fig. 1, where the exponent $L = 0$ when both momenta are simultaneously incoming or outgoing, and
$L = 1$ \mbox{otherwise}. The Feynman rules for the vertex with an arbitrary number of photons attached to the same blob will be given by the product of the corresponding number of single photon vertices.
\begin{figure}
\includegraphics[width=.45\textwidth]{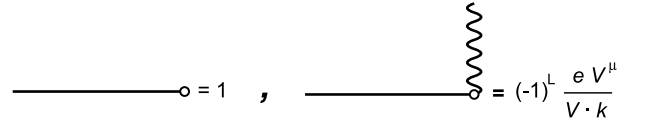}
\caption{Feynman Rules}
\label{fig:feynco}
\end{figure}

An arbitrary dressed state of two charged qubits would have the form
\begin{equation}\label{estadogeneral2}\ket\phi=\sum_{\sigma_1,\sigma_2}\int d^3p_1\,d^3p_2\, \varphi^d_{\sigma_1\sigma_2}(p_1,p_2)\,\ket{\sigma_1,p_1\,;\,\sigma_2,p_2}_d
\end{equation}
where the amplitudes
\begin{equation}\varphi^d_{\sigma_1\sigma_2}(p_1,p_2)=\,
 _d\braket{\sigma_1,p_1\,;\,\sigma_2,p_2}{\phi}
 \end{equation}
have to be computed taking into account the new vertices.  These amplitudes are infrared finite \cite{BaganPR}. The effect of the soft corrections in quantum information will finally show up in the entanglement measures for the bipartite system.

We show in Fig. 2  the one-loop corrections to the amplitude $\varphi$ due to virtual soft photons. In \cite{BaganPR} it is also shown that the summation of all the contributions at all orders in perturbation theory results in the factorized expression that exponentiates the one-loop results
\begin{equation}\label{damplitude}\varphi^d_{\sigma_1\sigma_2}(p_1,p_2)=e^{-ie^2
 \kappa_{vv'}}e^{Cvv'}G_v G_{v'}e^{D}\varphi_{\sigma_1,\sigma_2}(p_1,p_2)\end{equation}
The origins of the different correction factors are the following
\begin{itemize}
\item $e^{D}$ is the contribution of the standard QED IR-divergent diagrams (given at one-loop by the diagrams a, b and c).
\item $G_v$ and $G_{v'}$ are the contributions coming from the diagrams connecting a dressed vertex and a standard vertex (diagrams d and e at one-loop).
\item $e^{C_{vv'}}$ is the contribution of the diagrams with soft photons connecting both dressed vertices, (diagram f at one-loop).
\end{itemize}
\begin{figure}
\includegraphics[width=.40\textwidth]{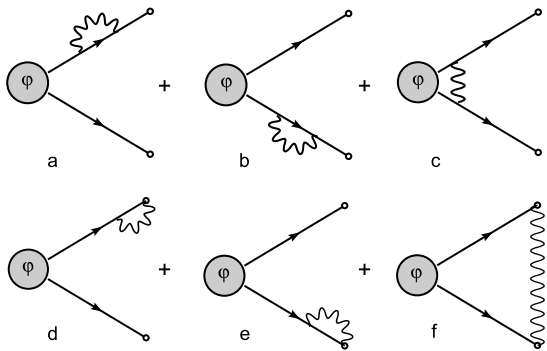}
\caption{One-loop virtual corrections to the amplitude}
\label{fig:feynco2}
\end{figure}
The factor $e^{-ie^2 \kappa_{vv'}}$ is a C-number that comes from the application of the dressing phase factor on our state, and it can be computed using the Hadamard Lemma.

Some of the diagrams of Fig. 2 are also UV divergent; wave function renormalization can be used to remove UV and IR divergencies altogether \cite{BaganPR,{Weinberg}}. The dressed wave function renormalization constant $Z_v$ can be given in terms of the standard renormalization constant $Z_2$ by
\begin{equation}\label{renorm}\sqrt{Z_v}=\sqrt{Z_2}G_v e^{-C_{vv}/2}\end{equation}
Taking this into account, we obtain the factor
\begin{equation}\label{damplitude2}\varphi^d_{\sigma_1\sigma_2}(p_1,p_2)=e^{-ie^2
\kappa_{vv'})}e^{C(p_1,p_2)+D(p_1,p_2)}
\varphi_{\sigma_1,\sigma_2}(p_1,p_2)\end{equation}
where a factor $Z_2^{-1}$ has been absorbed in the amplitude $\varphi$ in the rhs. Summarizing, $C(p_1,p_2)=C_{v_1v_2}+\frac{C_{v_1v_1} +C_{v_2v_2}}{2}$ are one-loop contributions from the dressing vertices, and $D(p_1,p_2)$ is the one-loop IR divergent contribution that comes from the standard QED diagrams a, b, and c of Fig. 2. The sum
\begin{equation}\label{F}F(p_1,p_2)=C(p_1,p_2) + D(p_1,p_2)\end{equation}
is IR finite \cite{BaganPR}.

The dressing phase factor is expected to cancel the phase divergences that appear in asymptotic evolution \eqref{4} \cite{Chargesingauge}, giving an overall finite phase $e^{ie^2\zeta(u_r)}$ depending on the relative velocity of the particles.

Taking everything into account we can finally obtain the correction for the amplitudes as
\begin{equation}\label{damplitude3}\varphi^d_{\sigma_1\sigma_2}(p_1,p_2)=e^{ie^2\zeta(u_r)}e^{F(p_1,p_2)}
\varphi_{\sigma_1,\sigma_2}(p_1,p_2).\end{equation}
A note of caution is that this is only possible when done on mass shell, i.e. when $p^\mu= m \gamma (\eta+v)^\mu$. With this caveat in mind, not only the probabilities, but also the very amplitudes result finite when we compute the corrections due to virtual soft photons.

Summing up, we have seen how to build physical qubits that are gauge invariant, asymptotically well-behaved and with the adequate pole structure associated with massive particles. As they also do not present IR divergences in their propagators, we argue that qubits built from dressed fields are the proper candidates for qubits in QED.

An illustrative example of the impact of those finite correction factors on the entanglement measures is given by A system of two charged spin $1/2$ particles, with spin components $\sigma_1$, $\sigma_2$, and relative velocity $\bm v =(v,\theta)$ in the CoM frame. Focus on its spin entanglement for which phase corrections will vanish, the density matrix can be written as
\begin{equation}\label{rho}\rho^\text{d}_{\sigma_{_1}\sigma_{_2}\sigma'_1\sigma'_2}(v,\theta)=
e^{2 F}\varphi_{\sigma_{_1}\sigma_{_2}}(v,\theta)\cdot
\varphi^\dagger_{\sigma'_1\sigma'_2}(v,\theta)\end{equation}

The reduced density matrix of particle 1, which is necessary for quantum information tasks, is given by
\begin{equation}\label{rho2}\rho^d_{\sigma_{_1}\sigma'_{_1}}(v,\theta) =e^{2 F(v)}
\rho_{\sigma_{_1}\sigma'_{_1}}(v,\theta)\end{equation}
where
\begin{equation}\label{rho3}\rho_{\sigma_{_1}\sigma'_{_1}}(v,\theta)=\sum_{\sigma}
\varphi_{\sigma_{_1}\sigma }(v,\theta)\cdot
\varphi^\dagger_{\sigma'_1\sigma }(v,\theta)\end{equation}
would be the density ignoring soft photon corrections. The entropy of entanglement $S^d=\mbox{Tr}(\rho^d \log \rho^d)$ of the dressed state is given by
\begin{equation}\label{Ent}S^d= e^{2F}(S+ 2F \mbox{Tr} \rho)\end{equation}
in terms of the entropy $S$ computed with $\rho$ discarding the correction. Finally, we should obtain
$e^{2F}(S+ 2F)$, or $e^{2F}S+ 2F$ for $S^d$ depending on which density, $\rho$ or $\rho^d$ respectively, would be normalized to one.

The specific computation of the finite correction factors on the entanglement measures coming from the IR structure of QED  is, as a matter of fact, a non trivial problem. There are two main different theoretical approaches to compute them, the Bloch-Nordsieck  \cite{BlochNord} and the Lee-Nauenberg \cite{LeeNau} methods. However, the consistence of these methods has been questioned in the case in which there are massless particles in both the initial and final states \cite{Lavellecol1,Lavellecol2}. On the experimental side, the IR factors are very difficult to measure since they are expected to be small and dependent upon the experiment resolution. Unfortunately, as far as the authors know, these finite correction factors have been poorly tested experimentally.

On the other hand, the lack of experimental evidence of these factors (like $F$ in our case) and the previous theoretical results \cite{Yennie1,GrammerYennie,Weinberg,BlochNord,LeeNau} indicates that they should be very small, so their impact on the entanglement measures would be negligible.

In conclusion, we have shown that, using existing techniques, it is possible to deal with charged qubits overcoming the troubles due to the masslessness of the photon. We have built qubits from the dressed fields which are gauge invariant and have a good asymptotic behavior. Also, their propagators have a mass shell pole and, more important, are IR finite. We have discussed the corrections introduced in the quantities of interest in quantum information in a particular case, showing that they are computable in terms of old known quantities and very small.

As a final remark, the finite phase factor $e^{ie^2\zeta}$ that is often neglected in the literature, (as it does not appear on probabilities) would be of interest in the cases where we consider entanglement involving superposition of states with different momenta. In these cases it would produce a relative phase that should be considered in quantum information tasks. This question should be addressed in the future.

The authors are indebted to E. Bagan, D. McMullan, and M. Lavelle for their invaluable help on the dressing procedure and infrared divergences. E.M-M thanks the hospitality at the School of Statistics and Mathematics of the University of Plymouth.

This work has been partially supported by the Spanish MEC Project FIS2005-05304. E. M-M is partially supported by the CSIC JAE-PREDOC2007 Grant and A1 mobility program.

\bibliographystyle{apsrev}

\begin{thebibliography}{28}
\expandafter\ifx\csname natexlab\endcsname\relax\def\natexlab#1{#1}\fi
\expandafter\ifx\csname bibnamefont\endcsname\relax
  \def\bibnamefont#1{#1}\fi
\expandafter\ifx\csname bibfnamefont\endcsname\relax
  \def\bibfnamefont#1{#1}\fi
\expandafter\ifx\csname citenamefont\endcsname\relax
  \def\citenamefont#1{#1}\fi
\expandafter\ifx\csname url\endcsname\relax
  \def\url#1{\texttt{#1}}\fi
\expandafter\ifx\csname urlprefix\endcsname\relax\def\urlprefix{URL }\fi
\providecommand{\bibinfo}[2]{#2}
\providecommand{\eprint}[2][]{\url{#2}}

\bibitem[{\citenamefont{Beckman et~al.}(2001)\citenamefont{Beckman, Gottesman,
  Nielsen, and Preskill}}]{Beckman}
\bibinfo{author}{\bibfnamefont{D.}~\bibnamefont{Beckman}},
  \bibinfo{author}{\bibfnamefont{D.}~\bibnamefont{Gottesman}},
  \bibinfo{author}{\bibfnamefont{M.~A.} \bibnamefont{Nielsen}},
  \bibnamefont{and} \bibinfo{author}{\bibfnamefont{J.}~\bibnamefont{Preskill}},
  \bibinfo{journal}{Phys. Rev. A} \textbf{\bibinfo{volume}{64}},
  \bibinfo{pages}{052309} (\bibinfo{year}{2001}).

\bibitem[{\citenamefont{Peres and Terno}(2003)}]{peresterno2}
\bibinfo{author}{\bibfnamefont{A.}~\bibnamefont{Peres}} \bibnamefont{and}
  \bibinfo{author}{\bibfnamefont{D.~R.} \bibnamefont{Terno}},
  \bibinfo{journal}{Int. J. Quant. Inf.} \textbf{\bibinfo{volume}{1}},
  \bibinfo{pages}{225} (\bibinfo{year}{2003}).

\bibitem[{\citenamefont{Peres and Terno}(2004)}]{RevPeresTerno}
\bibinfo{author}{\bibfnamefont{A.}~\bibnamefont{Peres}} \bibnamefont{and}
  \bibinfo{author}{\bibfnamefont{D.~R.} \bibnamefont{Terno}},
  \bibinfo{journal}{Rev. Mod. Phys.} \textbf{\bibinfo{volume}{76}},
  \bibinfo{pages}{93} (\bibinfo{year}{2004}).

\bibitem[{\citenamefont{Reznik et~al.}(2005)\citenamefont{Reznik, Retzker, and
  Silman}}]{reznik}
\bibinfo{author}{\bibfnamefont{B.}~\bibnamefont{Reznik}},
  \bibinfo{author}{\bibfnamefont{A.}~\bibnamefont{Retzker}}, \bibnamefont{and}
  \bibinfo{author}{\bibfnamefont{J.}~\bibnamefont{Silman}},
  \bibinfo{journal}{Phys. Rev. A} \textbf{\bibinfo{volume}{71}},
  \bibinfo{eid}{042104} (\bibinfo{year}{2005}).

\bibitem[{\citenamefont{Fuentes-Schuller and Mann}(2005)}]{fuentesschuller}
\bibinfo{author}{\bibfnamefont{I.}~\bibnamefont{Fuentes-Schuller}}
  \bibnamefont{and} \bibinfo{author}{\bibfnamefont{R.~B.} \bibnamefont{Mann}},
  \bibinfo{journal}{Phys. Rev. Lett.} \textbf{\bibinfo{volume}{95}},
  \bibinfo{pages}{120404} (\bibinfo{year}{2005}).

\bibitem[{\citenamefont{Verch}(2006)}]{Verch}
\bibinfo{author}{\bibfnamefont{R.}~\bibnamefont{Verch}},
  \bibinfo{journal}{Lect. Notes Phys.} \textbf{\bibinfo{volume}{702}},
  \bibinfo{pages}{133} (\bibinfo{year}{2006}).

\bibitem[{\citenamefont{Alsing et~al.}(2006)\citenamefont{Alsing,
  Fuentes-Schuller, Mann, and Tessier}}]{AlsingSchul}
\bibinfo{author}{\bibfnamefont{P.~M.} \bibnamefont{Alsing}},
  \bibinfo{author}{\bibfnamefont{I.}~\bibnamefont{Fuentes-Schuller}},
  \bibinfo{author}{\bibfnamefont{R.~B.} \bibnamefont{Mann}}, \bibnamefont{and}
  \bibinfo{author}{\bibfnamefont{T.~E.} \bibnamefont{Tessier}},
  \bibinfo{journal}{Phys. Rev. A} \textbf{\bibinfo{volume}{74}},
  \bibinfo{pages}{032326} (\bibinfo{year}{2006}).

\bibitem[{\citenamefont{Lamata et~al.}(2006)\citenamefont{Lamata,
  Martin-Delgado, and Solano}}]{LamataPRL}
\bibinfo{author}{\bibfnamefont{L.}~\bibnamefont{Lamata}},
  \bibinfo{author}{\bibfnamefont{M.~A.} \bibnamefont{Martin-Delgado}},
  \bibnamefont{and} \bibinfo{author}{\bibfnamefont{E.}~\bibnamefont{Solano}},
  \bibinfo{journal}{Physical Review Letters} \textbf{\bibinfo{volume}{97}},
  \bibinfo{pages}{250502} (\bibinfo{year}{2006}).

\bibitem[{\citenamefont{Caban and Rembielinski}(2006)}]{Caban}
\bibinfo{author}{\bibfnamefont{P.}~\bibnamefont{Caban}} \bibnamefont{and}
  \bibinfo{author}{\bibfnamefont{J.}~\bibnamefont{Rembielinski}},
  \bibinfo{journal}{Phys. Rev. A} \textbf{\bibinfo{volume}{74}},
  \bibinfo{pages}{042103} (\bibinfo{year}{2006}).

\bibitem[{\citenamefont{Silman and Reznik}(2007)}]{reznik2}
\bibinfo{author}{\bibfnamefont{J.}~\bibnamefont{Silman}} \bibnamefont{and}
  \bibinfo{author}{\bibfnamefont{B.}~\bibnamefont{Reznik}},
  \bibinfo{journal}{Phys. Rev. A} \textbf{\bibinfo{volume}{75}},
  \bibinfo{pages}{052307} (\bibinfo{year}{2007}).

\bibitem[{\citenamefont{Yeo et~al.}(2007)\citenamefont{Yeo, Lim, Ching, Chong,
  Chua, and Dewanto}}]{Chinos}
\bibinfo{author}{\bibfnamefont{Y.}~\bibnamefont{Yeo}},
  \bibinfo{author}{\bibfnamefont{Z.~H.} \bibnamefont{Lim}},
  \bibinfo{author}{\bibfnamefont{C.~L.} \bibnamefont{Ching}},
  \bibinfo{author}{\bibfnamefont{J.}~\bibnamefont{Chong}},
  \bibinfo{author}{\bibfnamefont{W.~K.} \bibnamefont{Chua}}, \bibnamefont{and}
  \bibinfo{author}{\bibfnamefont{A.}~\bibnamefont{Dewanto}},
  \bibinfo{journal}{Europhys. Lett.} \textbf{\bibinfo{volume}{78}},
  \bibinfo{pages}{20006} (\bibinfo{year}{2007}).

\bibitem[{\citenamefont{Leon and Sabin}(2008)}]{JleonSabin2a}
\bibinfo{author}{\bibfnamefont{J.}~\bibnamefont{Leon}} \bibnamefont{and}
  \bibinfo{author}{\bibfnamefont{C.}~\bibnamefont{Sabin}},
  \bibinfo{journal}{Phy. Rev. A} \textbf{\bibinfo{volume}{78}},
  \bibinfo{pages}{052314} (\bibinfo{year}{2008}).

\bibitem[{\citenamefont{Leon and Sabin}(2009)}]{JleonSabin2b}
\bibinfo{author}{\bibfnamefont{J.}~\bibnamefont{Leon}} \bibnamefont{and}
  \bibinfo{author}{\bibfnamefont{C.}~\bibnamefont{Sabin}},
  \bibinfo{journal}{Phy. Rev. A} \textbf{\bibinfo{volume}{79}},
  \bibinfo{pages}{012301} (\bibinfo{year}{2009}).

\bibitem[{\citenamefont{Kulish and Faddeev}(1970)}]{KulishFaddeev}
\bibinfo{author}{\bibfnamefont{P.}~\bibnamefont{Kulish}} \bibnamefont{and}
  \bibinfo{author}{\bibfnamefont{L.}~\bibnamefont{Faddeev}},
  \bibinfo{journal}{Theor. Mat. Fiz.} \textbf{\bibinfo{volume}{4}},
  \bibinfo{pages}{153} (\bibinfo{year}{1970}).

\bibitem[{\citenamefont{Jauch and Rohrlich}(1976)}]{JR}
\bibinfo{author}{\bibfnamefont{J.~M.} \bibnamefont{Jauch}} \bibnamefont{and}
  \bibinfo{author}{\bibfnamefont{F.}~\bibnamefont{Rohrlich}},
  \emph{\bibinfo{title}{The theory of photons and electrons}}
  (\bibinfo{publisher}{Springer-Verlag}, \bibinfo{address}{Berlin},
  \bibinfo{year}{1976}).

\bibitem[{\citenamefont{Bagan et~al.}(2000{\natexlab{a}})\citenamefont{Bagan,
  Lavelle, and McMullan}}]{BaganC}
\bibinfo{author}{\bibfnamefont{E.}~\bibnamefont{Bagan}},
  \bibinfo{author}{\bibfnamefont{M.}~\bibnamefont{Lavelle}}, \bibnamefont{and}
  \bibinfo{author}{\bibfnamefont{D.}~\bibnamefont{McMullan}},
  \bibinfo{journal}{Annals Phys.} \textbf{\bibinfo{volume}{282}},
  \bibinfo{pages}{471} (\bibinfo{year}{2000}{\natexlab{a}}).

\bibitem[{\citenamefont{Schroer}(1963)}]{Infraparticles}
\bibinfo{author}{\bibfnamefont{B.}~\bibnamefont{Schroer}},
  \bibinfo{journal}{Fortschr. Phys.} \textbf{\bibinfo{volume}{11}},
  \bibinfo{pages}{1} (\bibinfo{year}{1963}).

\bibitem[{\citenamefont{Bagan et~al.}(2000{\natexlab{b}})\citenamefont{Bagan,
  Lavelle, and McMullan}}]{BaganPR}
\bibinfo{author}{\bibfnamefont{E.}~\bibnamefont{Bagan}},
  \bibinfo{author}{\bibfnamefont{M.}~\bibnamefont{Lavelle}}, \bibnamefont{and}
  \bibinfo{author}{\bibfnamefont{D.}~\bibnamefont{McMullan}},
  \bibinfo{journal}{Annals Phys.} \textbf{\bibinfo{volume}{282}},
  \bibinfo{pages}{503} (\bibinfo{year}{2000}{\natexlab{b}}).

\bibitem[{\citenamefont{Chung}(1965)}]{Chung}
\bibinfo{author}{\bibfnamefont{V.}~\bibnamefont{Chung}},
  \bibinfo{journal}{Phys. Rev.} \textbf{\bibinfo{volume}{140}},
  \bibinfo{pages}{B1110} (\bibinfo{year}{1965}).

\bibitem[{\citenamefont{Georgi}(1990)}]{Georgi}
\bibinfo{author}{\bibfnamefont{H.}~\bibnamefont{Georgi}},
  \bibinfo{journal}{Phys. Lett. B} \textbf{\bibinfo{volume}{240}},
  \bibinfo{pages}{447} (\bibinfo{year}{1990}).

\bibitem[{\citenamefont{Weinberg}(1995)}]{Weinberg}
\bibinfo{author}{\bibfnamefont{S.}~\bibnamefont{Weinberg}},
  \emph{\bibinfo{title}{The Quantum Theory of Fields}},
  vol.~\bibinfo{volume}{1} (\bibinfo{publisher}{Cambrigde Univeristy Press},
  \bibinfo{address}{Cambrigde, England}, \bibinfo{year}{1995}).

\bibitem[{\citenamefont{Horan et~al.}(1998)\citenamefont{Horan, Lavelle, and
  McMullan}}]{Chargesingauge}
\bibinfo{author}{\bibfnamefont{R.}~\bibnamefont{Horan}},
  \bibinfo{author}{\bibfnamefont{M.}~\bibnamefont{Lavelle}}, \bibnamefont{and}
  \bibinfo{author}{\bibfnamefont{D.}~\bibnamefont{McMullan}},
  \bibinfo{journal}{Pramana, J. Phys.} \textbf{\bibinfo{volume}{51}},
  \bibinfo{pages}{317} (\bibinfo{year}{1998}).

\bibitem[{\citenamefont{Bloch and Nordsieck}(1937)}]{BlochNord}
\bibinfo{author}{\bibfnamefont{F.}~\bibnamefont{Bloch}} \bibnamefont{and}
  \bibinfo{author}{\bibfnamefont{A.}~\bibnamefont{Nordsieck}},
  \bibinfo{journal}{Phys. Rev.} \textbf{\bibinfo{volume}{52}},
  \bibinfo{pages}{54} (\bibinfo{year}{1937}).

\bibitem[{\citenamefont{Lee and Nauenberg}(1964)}]{LeeNau}
\bibinfo{author}{\bibfnamefont{T.~D.} \bibnamefont{Lee}} \bibnamefont{and}
  \bibinfo{author}{\bibfnamefont{M.}~\bibnamefont{Nauenberg}},
  \bibinfo{journal}{Phys. Rev.} \textbf{\bibinfo{volume}{133}},
  \bibinfo{pages}{B1549} (\bibinfo{year}{1964}).

\bibitem[{\citenamefont{Lavelle and McMullan}(2006)}]{Lavellecol1}
\bibinfo{author}{\bibfnamefont{M.}~\bibnamefont{Lavelle}} \bibnamefont{and}
  \bibinfo{author}{\bibfnamefont{D.}~\bibnamefont{McMullan}},
  \bibinfo{journal}{J. High Energy Phys.} \textbf{\bibinfo{volume}{2006}},
  \bibinfo{pages}{026} (\bibinfo{year}{2006}).

\bibitem[{\citenamefont{Lavelle and McMullan}(2007)}]{Lavellecol2}
\bibinfo{author}{\bibfnamefont{M.}~\bibnamefont{Lavelle}} \bibnamefont{and}
  \bibinfo{author}{\bibfnamefont{D.}~\bibnamefont{McMullan}},
  \bibinfo{journal}{Nucl. Phys. Proc. Suppl.} \textbf{\bibinfo{volume}{174}},
  \bibinfo{pages}{51} (\bibinfo{year}{2007}).

\bibitem[{\citenamefont{Yennie et~al.}(1961)\citenamefont{Yennie, Frautschi,
  and Suura}}]{Yennie1}
\bibinfo{author}{\bibfnamefont{D.~R.} \bibnamefont{Yennie}},
  \bibinfo{author}{\bibfnamefont{S.~C.} \bibnamefont{Frautschi}},
  \bibnamefont{and} \bibinfo{author}{\bibfnamefont{H.}~\bibnamefont{Suura}},
  \bibinfo{journal}{Ann. Phys.} \textbf{\bibinfo{volume}{13}},
  \bibinfo{pages}{379} (\bibinfo{year}{1961}).

\bibitem[{\citenamefont{Grammer and Yennie}(1973)}]{GrammerYennie}
\bibinfo{author}{\bibfnamefont{G.}~\bibnamefont{Grammer}} \bibnamefont{and}
  \bibinfo{author}{\bibfnamefont{D.~R.} \bibnamefont{Yennie}},
  \bibinfo{journal}{Phys. Rev. D} \textbf{\bibinfo{volume}{8}},
  \bibinfo{pages}{4332} (\bibinfo{year}{1973}).

\end{thebibliography}

\end{document}